\documentclass[nofootinbib,reprint,amsmath,amssymb,aps]{revtex4-1}
\usepackage{graphicx}
\usepackage[T1]{fontenc}
\usepackage{dcolumn}
\usepackage{bm}
\newcommand{\qm}[1]{``#1''}
\usepackage{hyperref}
\hypersetup{colorlinks, linkcolor={red},citecolor={blue},urlcolor={blue}}  
\newcommand\ChangeRT[1]{\noalign{\hrule height #1}}

\begin{document}

\preprint{APS/123-QED}

\title[Three-dimensional general relativistic Poynting-Robertson effect. IV]{Three-dimensional general relativistic Poynting-Robertson effect. IV.\\ Slowly rotating and non-spherical quadrupolar massive source}

\author{Vittorio De Falco$^{1}$}\email{vittorio.defalco@physics.cz}
\author{Maciek Wielgus$^{2,3}$}\email{maciek.wielgus@gmail.com}
\vspace{0.5cm}

\affiliation{$^1$ Department of Mathematics and Applications \qm{R. Caccioppoli}, University of Naples Federico II, Via Cintia, 80126 Naples, Italy\\
$^2$ Black Hole Initiative at Harvard University, 20 Garden Street, Cambridge, MA 02138, USA\\
$^3$ Center for Astrophysics  $|$ Harvard \& Smithsonian, 60 Garden Street, Cambridge, MA 02138, USA}

\date{\today}

\begin{abstract}
We consider a further extension of our previous works in the treatment of the three-dimensional general relativistic Poynting-Robertson effect, which describes the motion of a test particle around a compact object as affected by the radiation field originating from a rigidly rotating and spherical emitting source, which produces a radiation pressure, opposite to the gravitational pull, and a radiation drag force, which removes energy and angular momentum from the test particle. The gravitational source is modeled as a non-spherical and slowly rotating compact object endowed with a mass quadrupole moment and an angular momentum and it is formally described by the Hartle-Thorne metric. We derive the test particle's equations of motion in the three-dimensional and two-dimensional cases. We then investigate the properties of the critical hypersurfces (regions, where a balance between gravitational and radiation forces is established). Finally, we show how this model can be applied to treat radiation phenomena occurring in the vicinity of a neutron star. 
\end{abstract}

\maketitle
\section{Introduction}
\label{sec:intro}
The motion of matter around compact objects, when it is influenced by an electromagnetic radiation field (originating for example from the surface of a neutron star (NS), a boundary layer around a NS, a type-I X-ray burst on the NS polar caps, a hot corona around  a black hole (BH) or an accretion disk around a NS or a BH) deviates from a geodesic trajectory. In such processes, the general relativistic Poynting-Robertson (PR) effect plays an important role in removing energy and angular momentum from the affected body, thus playing a role of a dissipative force in General Relativity (GR) \cite{Poynting1903,Robertson1937}.  

Recently, a series of programmatic studies on such an effect in GR have been published. From a theoretical perspective, it is worth to cite: the general relativistic modeling from the two-dimensional (2D) \cite{Bini2009,Bini2011,Bini2011v,Bini2015} to the 3D cases in Kerr and other metrics \cite{DeFalco20183D,Bakala2019,Wielgus2019,Defalco2020III}, its treatment under a Lagrangian formalism, determining for the first time the analytical form of the Rayleigh potential in GR literature \cite{DeFalco2018,DeFalco2019L,DeFalco2019VE}, proof that the equatorial ring of the critical hypersurface is a stable attractor \cite{Stahl2012}, and the whole critical hypersurface is a basin of attraction \cite{DeFalco2019ST}. There are also several attempts to apply such effect to describe astrophysical phenomena, like: analysis of the disk dynamical evolution when it is intercepted by a type-I X-ray burst \cite{Walker1989,Walker1992,Lancova2017,Fragile2020}, modeling the photospheric expansion occuring during Eddington-luminosity X-ray bursts \cite{Wielgus2015,Wielgus2016p} and associated oscillations \cite{Wielgus2012,Bollimpalli2019}, a new method to diagnose the presence of wormholes through the detection of metric-changes occurring in strong field regimes around black holes through the PR critical hypersurfaces \cite{Defalco2020wh}.

The exterior spacetime of rotating NS (or other compact objects such as dark stars, gravastars, boson stars \cite{Cardoso2019}) is not unique and analytically known. However, if we consider such objects in a regime of slow rigid rotation, modeled as stationary and axially symmetric perfect fluids through mass $M$, angular momentum $J$, and quadrupole moment $Q$, then the \emph{Hartle-Thorne metric can be employed to realistically describe them with accuracy up to the second order in $J$, and first order in $Q$} \cite{Hartle1967,Hartle1968}. While Schwarzschild, Kerr, and Erez-Rosen metrics are all exact solutions of Einstein vacuum field equations to respectively model static, rotating, and static, axially symmetric and non-spherical compact objects, \emph{the Hartle-Thorne metric, on the other hand, is an approximate solution of Einstein field equations in the vacuum}.

In this work, we aim to extend our previous works on the 3D modeling of the general relativistic PR effect \cite{DeFalco20183D,Bakala2019,Defalco2020III} in the Hartle-Thorne metric, using as a description of the radiation field the model developed in Ref. \cite{Bakala2019}. The paper is organized as follows: in Sec. \ref{sec:geometry} we recall the Hartle-Thorne metric; in Sec. \ref{sec:dynamics} we derive the test particle's equations of motion; in Sec. \ref{sec:CH} we analyse the critical hypersurfaces, investigating extensively their properties, and showing also a possibile application to the NS case; finally in Sec. \ref{sec:end} we draw our conclusions. 

\section{Spacetime geometry}
\label{sec:geometry}

\subsection{Hartle-Thorne spacetime}
\label{sec:HT_met}
Astrophysical objects are not exactly spherical symmetric, the more if they are rotating. The exterior spacetime of a slowly rotating and slightly deformed compact object endowed with total mass $M$, angular momentum $J$ and quadrupole parameter $Q$ \footnote{The quadrupole parameter $Q$ introduced here is related to the mass quadrupole moment $Q_{HT}$ as defined by Hartle and Thorne \cite{Hartle1967,Hartle1968} through the formula $Q=2J^2/M-Q_{HT}$.} can be accurately described by the Hartle-Thorne metric \cite{Hartle1967,Hartle1968}. Using geometrical units, where $G=c=1$, and spherical coordinates $(t,r,\theta,\varphi)$, the line element of this metric is given by \cite{Bini2013n}
\begin{eqnarray}\label{eq:HTmetric}
ds^2&=&-f(r)F_1(r,\theta)dt^2+\frac{F_2(r,\theta)}{f(r)}dr^2\\\nonumber
&+&r^2F_3(r,\theta)(d\theta^2+\sin^2\theta d\varphi^2)-\frac{4J}{r}\sin^2\theta\ dt\ d\varphi,
\end{eqnarray}
where
\begin{eqnarray}\label{eq:ht1}
f(r)&=&1-\frac{2{ M }}{r},\notag\\
k_1(r)&=&\frac{J^{2}}{{M}r^3}f(-2r)-\frac{5}{8}\frac{Q-J^{2}/{M}}{{M}^3}Q_2\left(\frac{r}{{M}}-1\right),\notag\\
k_2(r)&=&k_1(r)-\frac{6J^{2}}{r^4}, \nonumber\\
k_3(r)&=&k_1(r)+\frac{J^{2}}{r^4}-\frac{5}{4}\frac{Q-J^{2}/{M}}{{M}^2r\sqrt{f(r)}}Q_1\left(\frac{r}{M}-1\right),\notag\\
F_1(r,\theta)&=&1+2k_1(r)P_2(\cos\theta)+\frac{2}{f(r)}\frac{J^{2}}{r^{4}}(2\cos^2\theta-1),\notag\\
F_2(r,\theta)&=&1-2k_2(r)P_2(\cos\theta)-\frac{2}{f(r)}\frac{J^{2}}{r^4},\notag\\
F_3(r,\theta)&=&1-2k_3(r)P_2(\cos\theta),
\end{eqnarray}
and $P_{2}(x)$ is the Legendre polynomials of the first kind, $Q_1(x),Q_2(x)$ are the associated Legendre polynomials of the second kind, which all explicitly read as
\begin{eqnarray}\label{eq:ht2}
P_2(x)&=&\frac{1}{2}(3x^{2}-1),\nonumber\\
Q_1(x)&=&(x^{2}-1)^{1/2}\left[\frac{3x}{2}\ln\left(\frac{x+1}{x-1}\right)-\frac{3x^{2}-2}{x^{2}-1}\right],\\
Q_2(x)&=&(x^{2}-1)\left[\frac{3}{2}\ln\left(\frac{x+1}{x-1}\right)-\frac{3x^{3}-5x}{(x^{2}-1)^2}\right].\nonumber
\end{eqnarray}

\subsubsection{Properties of the Hartle-Thorne metric}
We report some useful properties of the Hartle-Thorne metric, which will be useful in the next sections.
\begin{itemize}
\item Metric (\ref{eq:HTmetric}) reduces to the approximate Kerr metric in the Boyer-Lindquist coordinates $(t,\ R,\ \Theta,\ \varphi)$ up to second order terms in the rotation parameter $a$ by considering $J=-M^2a$, $Q=J^2/M$, and the following transformation of coordinates \cite{Bini2013n}
\begin{eqnarray}\label{tr2}
r&=&R+\frac{a^2}{2R}\bigg[f(-R)f(2R)\notag\\
&&- f(R)f\left(-\frac{3}{2}R\right)\cos^2\Theta\bigg], \\
\theta&=&\Theta+\frac{a^2}{2R^2}f(-R)\sin\Theta\cos\Theta.\notag
\end{eqnarray}
\item For $J=0$ metric (\ref{eq:HTmetric}) reduces to the linearized Erez-Rosen spacetime (static, axially symmetric, and non-spherical quadrupolar massive source) with respect to its quadrupole parameter $Q_{ER}$ \cite{Bini2015}. 
\item Hartle-Thorne spacetime can admit an event horizon $R_H$ and ergosphere $R_{SL}$, whose expressions are obtained by respectively imposing $g_{t\varphi}^2-g_{tt}g_{\varphi\varphi}=0$, and $g_{tt}=0$, and then solving such equations for $r$ in terms of $\theta$, once $Q,J$ has been assigned \cite{Abramowicz2003}.
\item The domain of validity of the Hartle-Thorne approximation around a gravitating body must always be (see Sec. 2 in Ref. \cite{Bini2013n}, for further details)
\begin{equation} \label{eq:HTvalidity}
r-2M\gg \left(\frac{25J^2Q^2}{128M^7}\right)^{1/3}.
\end{equation}
\end{itemize}

\subsection{Zero angular momentum observers}
\label{sec:obsrever}
The Hartle-Thorne spacetime admits, as in the Kerr metric, zero angular momentum observers (ZAMOs), who are dragged by the rotation of the spacetime (even though it is not strong) with angular velocity $\Omega_{\mathrm{ZAMO}}=-g_{\varphi t}/g_{\varphi \varphi}$, while their radial and latitudinal coordinates remain constant. The four-velocity of ZAMOs, $\boldsymbol{n}$, is \cite{Bini2009,Bini2011,DeFalco20183D,Bakala2019},
\begin{equation}
\label{n}
\boldsymbol{n}=\frac{1}{N}(\boldsymbol{\partial_t}-N^{\varphi}\boldsymbol{\partial_\varphi})\,,
\end{equation}
where $N=(-g^{tt})^{-1/2}$ is the time lapse function, and $N^{\varphi}=g_{t\varphi}/g_{\varphi\varphi}$ is the spatial shift vector field. An orthonormal frame adapted to the ZAMOs is \cite{Bini2009,Bini2011,DeFalco20183D,Bakala2019}
\begin{equation}\label{eq:zamoframe}
\begin{aligned}
&\boldsymbol{e_{\hat t}}=\boldsymbol{n},\
\boldsymbol{e_{\hat r}}=\frac{\boldsymbol{\partial_r}}{\sqrt{g_{rr}}},\
\boldsymbol{e_{\hat \theta}}=\frac{\boldsymbol{\partial_\theta}}{\sqrt{g_{\theta\theta}}},\
\boldsymbol{e_{\hat \varphi}}=\frac{\boldsymbol{\partial_\varphi}}{\sqrt{g_{\varphi\varphi}}}.
\end{aligned}
\end{equation}
All the indices associated to the ZAMO frame will be labeled by a hat, instead all the quantities measured in the ZAMO frame will be followed by $(\boldsymbol{n})$. 

\subsection{ZAMO kinematical quantities}
\label{sec:ZAMOq}
The properties of the Hartle-Thorne spacetime combines those of the Kerr \cite{DeFalco20183D,Bakala2019} and Erez-Rosen metrics \cite{Defalco2020III}. Therefore, as done in the previous cases it is still convenient to use the Lie transport (see \cite{Bini1997a,Bini1997b,DeFalco2018}, for further details), where the nonzero ZAMO kinematical quantities are: acceleration $\boldsymbol{a}(\boldsymbol{n})=\nabla_{\boldsymbol{n}} \boldsymbol{n}$, expansion tensor along the $\hat{\varphi}$-direction $\boldsymbol{\theta_{\hat\varphi}}(\boldsymbol{n})$, and the signed Lie curvature tensors $\boldsymbol{k}(x^i,\boldsymbol{n})$ relative to the ZAMO $\boldsymbol{n}$ four-velocity along the directions $x^i=r,\theta,\varphi$ \cite{Bini1997a,Bini1997b,DeFalco20183D}. They have only nonzero components in the $\boldsymbol{\hat{r}}-\boldsymbol{\hat{\theta}}$ ZAMO plane \cite{DeFalco20183D,Bakala2019,Defalco2020III}, and can be calculated through the formulas:
\begin{equation}\label{eq:kinematics}
\begin{aligned}
\boldsymbol{a}(\boldsymbol{n})&=a(\boldsymbol{n})^{\hat r}\ \boldsymbol{e_{\hat r}}+a(\boldsymbol{n})^{\hat \theta}\ \boldsymbol{e_{\hat \theta}}\\
&=\frac{\partial_r N}{N\sqrt{g_{rr}}}\ \boldsymbol{\partial_{r}}+\frac{\partial_\theta N}{N\sqrt{g_{\theta\theta}}}\ \boldsymbol{\partial_{\theta}},\\
\boldsymbol{\theta_{\hat\varphi}}(\boldsymbol{n})& = \theta(\boldsymbol{n})^{\hat r}{}_{\hat\varphi}\, \boldsymbol{e_{\hat r}} + \theta(\boldsymbol{n})^{\hat\theta}{} _{\hat\varphi}\, \boldsymbol{e_{\hat \theta}}\\
&= -\frac{\sqrt{g_{\varphi\varphi}}}{2N}\,\left(\frac{\partial_{r} N^\varphi}{\sqrt{g_{rr}}}\, \boldsymbol{\partial_{r}} + \frac{\partial_{\theta} N^\varphi}{\sqrt{g_{\theta\theta}}} \boldsymbol{\partial_{\theta}}\right),\\
\boldsymbol{k}(x^i,\boldsymbol{n})&=k(x^i,\boldsymbol{n})^{\hat r}\ \boldsymbol{e_{\hat r}}+k(x^i,\boldsymbol{n})^{\hat \theta}\ \boldsymbol{e_{\hat \theta}}\\
&=-\frac{\partial_r g_{ii}}{2g_{ii}\sqrt{g_{rr}}}\ \boldsymbol{\partial_{r}}-\frac{\partial_\theta g_{ii}}{2g_{ii}\sqrt{g_{\theta\theta}}}\ \boldsymbol{\partial_{\theta}},
\end{aligned}
\end{equation}
It is important to note that $\boldsymbol{k}(\varphi,\boldsymbol{n})=\boldsymbol{k}_{\rm (Lie)}(\boldsymbol{n})$ \cite{DeFalco20183D,Defalco2020III}. The ZAMO kinematical quantities are expressed in terms of the derivatives of $\partial_\alpha N, \partial_\alpha N^\varphi$, where $\alpha=r,\theta$, namely
\begin{eqnarray}
\psi_\alpha&=&g_{t\varphi}^2\partial_\alpha g_{\varphi\varphi}+g_{\varphi\varphi}(g_{\varphi\varphi}\partial_\alpha g_{tt}- 2g_{t\varphi}\partial_\alpha g_{t\varphi}),\notag\\
\partial_\alpha N&=&-\frac{N^3}{2}\frac{\psi_\alpha}{(g_{t\varphi}^2-g_{tt}g_{\varphi\varphi})^2},\\
\partial_\alpha N^\varphi&=&\frac{g_{\varphi\varphi}\partial_\alpha g_{t\varphi}-g_{t\varphi}\partial_\alpha g_{\varphi\varphi}}{g_{\varphi\varphi}^2}.\notag
\end{eqnarray}
Therefore, to have theirs explicit expressions we need to calculate the derivatives of the metric components with respect to the radial $r$ and polar $\theta$ coordinates. The derivatives with respect to $r$ are 
\begin{eqnarray}
&&\partial_r g_{tt}=-[\partial_r f(r)F_1(r,\theta)+f(r) \partial_rF_1(r,\theta)],\notag\\
&&\partial_r g_{rr}=\frac{f(r) \partial_rF_2(r,\theta)-\partial_r f(r)F_2(r,\theta)}{f(r)^2},\notag\\
&&\partial_r g_{\theta\theta}=r[2F_3+r\partial_rF_3(r,\theta)],\\
&&\partial_r g_{\varphi\varphi}=\partial_r g_{\theta\theta} \sin^2\theta,\notag\\
&&\partial_r g_{t\varphi}=\frac{2J}{r^2}\sin^2\theta,\notag
\end{eqnarray}
while the derivatives with respect to $\theta$ are
\begin{eqnarray}
&&\partial_\theta g_{tt}=-f(r)\partial_\theta F_1(r,\theta),\notag\\
&&\partial_\theta g_{rr}=\frac{\partial_\theta F_2(r,\theta)}{f(r)},\notag\\
&&\partial_\theta g_{\theta\theta}=r^2\partial_\theta F_3(r,\theta),\\
&&\partial_\theta g_{\varphi\varphi}=r^2[\sin(2\theta)F_3(r,\theta)+\sin^2\theta\partial_\theta F_3(r,\theta)],\notag\\
&&\partial_\theta g_{t\varphi}=-\frac{2J}{r}\sin(2\theta).\notag
\end{eqnarray}
In Table \ref{tab:ZAMOq} we summarize the explicit expressions of functions' derivatives (\ref{eq:ht1}) and (\ref{eq:ht2}) in Hartle-Thorne metric. 

\renewcommand{\arraystretch}{1.8}
\begin{table*}[t!]
\begin{center}
\caption{\label{tab:ZAMOq} Explicit expressions of functions' derivatives (\ref{eq:ht1}) and (\ref{eq:ht2}) in Hartle-Thorne metric. For easing the notations, we define $x=r/M-1$ and $y=\cos\theta$, where $\partial_yP_2(y)=3y$ and $\partial_rf(r)=2M/r^2$.\\}	
\normalsize
\begin{tabular}{l  c} 
\ChangeRT{1pt}
{\bf Metric quantity} & {\bf Explicit expression} \\
\ChangeRT{1pt}
$\partial_x Q_1$ & $\frac{x(7-6x^2)+3(x^2-1)(x^2-1/2)\ln[(x+1)/(x-1)]}{(x^2-1)^{3/2}}$\\
\hline
$\partial_x Q_2$ & $\frac{-8+10x^2-6x^4+3x(x^2-1)^2\ln[(x+1)/(x-1)]}{(x^2-1)^2}$\\
\hline
\hline
$\partial_r k_1(r)$ &$-\frac{J^{2}}{{M}r^5}(3r+4M)-\frac{5}{8}\frac{Q-J^{2}/{M}}{{M}^4}\partial_xQ_2(x)$\\
\hline
$\partial_r k_2(r)$ & $\partial_r k_1(r)+\frac{24J^{2}}{r^5}$\\
\hline
$\partial_r k_3(r)$ & $\partial_r k_1(r)-\frac{4J^{2}}{r^5}-\frac{5}{4}\frac{Q-J^{2}/{M}}{{M}^2}\left[\frac{rf(r)\partial_x Q_1(x)/M- Q_1(x)f(2r)}{r^2\sqrt{f(r)^3}}\right]$\\
\hline
\hline
$\partial_r F_1(r,\theta)$ & $2\partial_r k_1(r)P_2(y)-\frac{4J^2(2\cos^2\theta-1)}{r^6f(r)^2}(2r-3M)$\\
\hline
$\partial_r F_2(r,\theta)$ & $-2\partial_r k_2(r)P_2(y)+\frac{4J^2}{r^6f(r)^2}(2r-3M)$\\
\hline
$\partial_r F_3(r,\theta)$ & $-2\partial_r k_3(r)P_2(y)$\\
\hline
\hline
$\partial_\theta F_1(r,\theta)$ & $-\sin(2\theta)\left[3k_1(r)+\frac{4J^2}{r^4f(r)}\right]$\\
\hline
$\partial_\theta F_2(r,\theta)$ & $3\sin(2\theta)k_2(r)$\\
\hline
$\partial_\theta F_3(r,\theta)$ & $3\sin(2\theta)k_3(r)$\\
\ChangeRT{1pt}
\end{tabular}
\end{center}
\end{table*}

\section{Test particle dynamics}
\label{sec:dynamics}

\subsection{Radiation field}
\label{sec:radfield}
In this section we approximate the radiation field by considering it to only consist locally of photons traveling on trajectories orthogonal to the rotating emission sphere (corresponding either to the NS surface or a boundary layer forming around a NS). Such photons are characterized by a four-momentum component $k^\theta = 0$. This approach follows previous studies \cite{Bini2009,Bini2011,DeFalco20183D,Bakala2019,Defalco2020III}, and it simplifies the treatment of the model significantly. Of course more astrophysically realistic models should take into account the photon emission from the whole surface including the whole range of outgoing light ray directions, the angular dependence of the surface emissivity, and law of emission related to the equation of state of the emitting surface. Although in the literature there are some attempts along this direction \cite{Abramowicz1990,Miller1996,Wielgus2019}, they are based on relativistic models simpler than the one proposed in this paper. These crucial features are not discussed in the present article, but they will be part of forthcoming works.

The effective description of the radiation field is thus given by the stress-energy tensor \cite{Bini2009,Bini2011,DeFalco20183D,Bakala2019,Defalco2020III}
\begin{equation} \label{eq:radfield}
T^{\alpha\beta}=\Phi^2 k^\alpha k^\beta,\quad k^\alpha k_\alpha=0,\quad k^\beta\nabla_\beta k^\alpha=0,
\end{equation}
where $\Phi$ is the parameter related to the intensity of the radiation field. The photon four-momentum $\boldsymbol{k}$ can be split in the ZAMO frame as \cite{Bini2009,Bini2011,DeFalco20183D,Bakala2019,Defalco2020III}
\begin{eqnarray} 
&&\boldsymbol{k}=E(\boldsymbol{n})[\boldsymbol{n}+\boldsymbol{\hat{\nu}} (\boldsymbol{k},\boldsymbol{n})],\label{eq:ksplit}\\
&&\boldsymbol{\hat{\nu}}(\boldsymbol{k},\boldsymbol{n})=\sin\beta\sin\xi\ \boldsymbol{e_{\hat r}}+\cos\xi\ \boldsymbol{e_{\hat \theta}}+\sin\xi\cos\beta\ \boldsymbol{e_{\hat \varphi}},\notag
\end{eqnarray}
where $\boldsymbol{\hat{\nu}} (\boldsymbol{k},\boldsymbol{n})$ is the photon spatial velocity on the spatial hypersurface orthogonal to $\boldsymbol{n}$, and $E(\boldsymbol{n})$ is the relative photon energy in the ZAMO frame \cite{Bini2009,Bini2011,DeFalco20183D,Bakala2019,Defalco2020III}
\begin{equation} \label{eq:ene}
E(\boldsymbol{n})=-\boldsymbol{k}\cdot\boldsymbol{n}=\frac{E}{N}(1+bN^\varphi),
\end{equation}
where $E=-k_t>0$ is the conserved photon energy, $\beta$ and $\xi$ are the two angles in the azimuthal and polar direction, respectively. The case $\sin\beta > 0$ corresponds to outgoing photons (increasing radial distance from the central source), and $\sin\beta < 0$ to incoming photons (decreasing $r$). The angular momentum along the polar $\boldsymbol{\hat \theta}$-axis in the local static observer frame, $L_{\hat \theta}(\boldsymbol{n})$ is \cite{DeFalco20183D,Bakala2019,Defalco2020III}
\begin{equation} \label{eq:angmom}
\begin{aligned}
E(\boldsymbol{n})\cos\beta\sin\xi&=L_{\hat \theta}(\boldsymbol{n})=
\boldsymbol{k}(\boldsymbol{n})\cdot \boldsymbol{e_{\hat \varphi}}=\frac{L_z}{\sqrt{g_{\varphi\varphi}}},
\end{aligned}
\end{equation}
where $L_z = k_\varphi$ is the conserved photon angular momentum along the $\boldsymbol{\theta}$-axis. From Eqs. (\ref{eq:ene}) and (\ref{eq:angmom}), we have  
\begin{equation} \label{eq:ang}
\cos\beta=\frac{bN}{\sqrt{g_{\varphi\varphi}}(1+bN^\varphi)},
\end{equation}
where $b = L_z/E$ denotes the azimuthal photon impact parameter associated to the azimuthal angle $\beta$. Following the same strategy as adopted in previous studies (see Refs. \cite{DeFalco20183D,Bakala2019,Defalco2020III}, for details) we assume that $k^\theta=0$ along all the photon trajectories. This implies that the polar angle $\theta$ is conserved along the photon trajectories, namely $\dot{\theta}=k^\theta/g_{\theta\theta}=0$, and so $\theta=const$. Since our radiation field is emitted radially in the frame of the rigidly rotating emitting surface, we have that $\xi=\pi/2$ (cf. Eq. (\ref{eq:ksplit})) and everything is expressed only in terms of the parameter $b$ and the angle $\theta$ occupied by the test particle. This implies that \cite{Bakala2019,Defalco2020III}
\begin{equation} \label{eq:imp_para}
b=\left[-\frac{g_{t\varphi}+g_{\varphi\varphi}\Omega_\star}{g_{tt}+g_{t\varphi}\Omega_\star}\right]_{r=R_\star},
\end{equation}
where $R_\star$ and $\Omega_\star$ are respectively radius and angular velocity of the emitting surface. 

Therefore, the photon four-momentum $\boldsymbol{k}$ is defined in terms of $b(\theta)$ or equivalently $(\theta,R_\star,\Omega_\star)$, whereas the stress-energy tensor of the radiation field (\ref{eq:radfield}) is completely determined by calculating the quantity $\Phi$. From the conservation equations $\nabla_{\beta}T^{\alpha\beta}=0$, the absence of photon latitudinal motion ($k^{\theta}=0$), and the axial symmetries of the Hartle-Thorne spacetime, we have \cite{Bini2009,Bini2011,DeFalco20183D,Bakala2019,Defalco2020III}
\begin{equation}\label{flux_cons}
0=\nabla_\beta (\Phi^2 k^\beta)=\partial_r (\sqrt{-g}\Phi^2 k^r).
\end{equation}
Therefore, we obtain \cite{Bini2011,DeFalco20183D,Defalco2020III}
\begin{equation} \label{eq:phi}
\sqrt{-g}\Phi^2 k^r\equiv N E(\boldsymbol{n})\sqrt{g_{\varphi\varphi}g_{\theta\theta}}\sin\beta\Phi^2=E\sin\theta\Phi_0^2,
\end{equation}
where $\Phi_0$ is $\Phi$ evaluated at the emitting surface. Then, after some algebra, we obtain
\begin{equation}\label{INT_PAR}
\Phi^2=\frac{\Phi_0^2}{(1+bN^\varphi)r^2F_3(r,\theta)\sin\beta}.
\end{equation}

\subsection{Test particle motion}
\label{sec:tpmotion}
A test particle moves in the 3D space with four-velocity $\boldsymbol{U}$ and spatial velocity $\boldsymbol{\hat{\nu}}(\boldsymbol{U},\boldsymbol{n})$ with respect to the ZAMO frame, given respectively by \cite{Bini2009,DeFalco20183D}
\begin{eqnarray} \label{eq:TPvelocity}
\boldsymbol{U}&=&\gamma(\boldsymbol{U},\boldsymbol{n})[\boldsymbol{n}+\boldsymbol{\nu} (\boldsymbol{U},\boldsymbol{n})],\\
\boldsymbol{\hat{\nu}}(\boldsymbol{U},\boldsymbol{n})&=&\nu^{\hat r}\boldsymbol{e_{\hat r}}+\nu^{\hat \theta}\boldsymbol{e_{\hat \theta}}+\nu^{\hat \varphi}\boldsymbol{e_{\hat \varphi}}\\
&=&\nu(\sin\alpha\sin\psi\ \boldsymbol{e_{\hat r}}+\cos\psi\ \boldsymbol{e_{\hat \theta}}+\sin\psi\cos\alpha\ \boldsymbol{e_{\hat \varphi}}),\notag
\end{eqnarray}
where $\gamma(\boldsymbol{U},\boldsymbol{n})\equiv \gamma=1/\sqrt{1-||\boldsymbol{\nu}(\boldsymbol{U},\boldsymbol{n})||^2}$ is the Lorentz factor, $\nu^{\hat\alpha}(\boldsymbol{U},\boldsymbol{n})\equiv \nu^{\hat\alpha}$ is the spatial velocity in the ZAMO frame, $\alpha$ and $\psi$ are the azimuthal and polar angles, respectively, and $\nu=\sqrt{\nu^{\hat r}{}^2+\nu^{\hat \theta}{}^2+\nu^{\hat \varphi}{}^2}$ is the module of the spatial velocity. The explicit expression of the test particle velocity components are
\begin{equation} \label{eq:velocitycomp}
\begin{aligned}
&U^{\hat t}\equiv \frac{dt}{d\tau}=\frac{\gamma}{N},\quad U^{\hat r}\equiv \frac{dr}{d\tau}=\frac{\gamma\nu^{\hat r}}{\sqrt{g_{rr}}},\\
&U^{\hat \theta}\equiv \frac{d\theta}{d\tau}=\frac{\gamma\nu^{\hat \theta}}{\sqrt{g_{\theta\theta}}},\quad U^{\hat \varphi}\equiv \frac{d\varphi}{d\tau}=\frac{\gamma\nu^{\hat \varphi}}{\sqrt{g_{\varphi\varphi}}}-\gamma \frac{N^\varphi}{N},
\end{aligned}
\end{equation}
where $\tau$ is the affine (or proper time) parameter along the test particle's world line.

Using the \emph{observer splitting formalism}, we find that the test particle acceleration in the Hartle-Thorne spacetime is similar to the that of the Erez-Rosen metric \cite{Defalco2020III}, whose explicit expression is given by:
\begin{eqnarray}
a(\boldsymbol{U})^{\hat r}&=& \gamma^2 \left[a(\boldsymbol{n})^{\hat r}+2\nu\cos\alpha\sin\psi\theta(\boldsymbol{n})^{\hat r}{}_{\hat \varphi}\right.\label{acc1} \\
&&\left.+\nu^2\left(-k(r,\boldsymbol{n})^{\hat \theta}\sin\alpha\sin\psi\cos\psi\right.\right.\notag\\
&&\left.\left.+k(\varphi,\boldsymbol{n})^{\hat r}\sin^2\psi\cos^2\alpha+k(\theta,\boldsymbol{n})^{\hat r}\cos^2\psi\right)\right]\nonumber\\
&&+\gamma \left(\gamma^2 \sin\alpha\sin\psi \frac{\rm d \nu}{\rm d \tau}+\nu \cos \alpha\sin\psi \frac{\rm d \alpha}{\rm d \tau}\right.\nonumber\\
&&\left.+\nu \cos \psi\sin\alpha \frac{\rm d \psi}{\rm d \tau} \right), \nonumber\\   
a(\boldsymbol{U})^{\hat \theta}&=&\gamma^2 \left[a(\boldsymbol{n})^{\hat \theta}+2\nu\cos\alpha\sin\psi\theta(\boldsymbol{n})^{\hat \theta}{}_{\hat \varphi}\right.\\
&&\left.+\nu^2\left(k(\varphi,\boldsymbol{n})^{\hat \theta}\cos^2\alpha\sin^2\psi\right.\right.\label{acc3}\\
&&\left.\left.+k(r,\boldsymbol{n})^{\hat \theta}\sin^2\alpha\sin^2\psi\right.\right.\nonumber\\
&&\left.\left.-k(\theta,\boldsymbol{n})^{\hat r}\sin\alpha\sin\psi\cos\psi\right)\right]\nonumber\\
&&+ \gamma\left(\gamma^2 \cos\psi \frac{\rm d \nu}{\rm d \tau}-\nu \sin\psi \frac{\rm d \psi}{\rm d \tau}\right).\nonumber\\
a(\boldsymbol{U})^{\hat \varphi}&=&-\gamma^2 \nu^2\left[k(\varphi,\boldsymbol{n})^{\hat \theta}\sin\psi\cos\alpha\cos\psi\right.\label{acc2}\\ 
&&\left.+k(\varphi,\boldsymbol{n})^{\hat r}\sin^2\psi\sin\alpha\cos\alpha\right]\nonumber\\
&&+ \gamma\left(\gamma^2 \cos \alpha\sin\psi \frac{\rm d \nu}{\rm d \tau}\right.\nonumber\\
&&\left.-\nu\sin \alpha\sin\psi \frac{\rm d \alpha}{\rm d \tau}+\nu\cos\alpha\cos\psi \frac{\rm d \psi}{\rm d \tau}\right).\nonumber
\end{eqnarray}
From the orthogonality between $\boldsymbol{a}(\boldsymbol{U})$ and $\boldsymbol{U}$, we can determine the expression of $a(\boldsymbol{U})^{\hat t}$ \cite{DeFalco20183D,Bakala2019}
\begin{eqnarray}\label{acc4}
a(\boldsymbol{U})^{\hat t}&=&\nu[a(\boldsymbol{U})^{\hat r}\sin\alpha\sin\psi+a(\boldsymbol{U})^{\hat \theta}\cos\psi\\
&&+a(\boldsymbol{U})^{\hat \varphi}\cos\alpha\sin\psi]\notag\\
&&=\gamma^2\nu\left[\sin\alpha\sin\psi\left(a(\boldsymbol{n})^{\hat r}+2\nu\cos\alpha\sin\psi\theta(\boldsymbol{n})^{\hat r}{}_{\hat \varphi}\right)\right.\notag\\
&&\left.+\cos\psi\left(a(\boldsymbol{n})^{\hat \theta}+2\nu\cos\alpha\sin\psi\theta(\boldsymbol{n})^{\hat \theta}{}_{\hat \varphi}\right)\right]+ \gamma^3\nu \frac{\rm d \nu}{\rm d \tau}.\notag
\end{eqnarray}
We note that such expressions for $J=0$ ($Q=0$) behave similarly to that of the Erez-Rosen \cite{Defalco2020III} (Kerr \cite{DeFalco20183D,Bakala2019}) metric. Instead, for $J=0,Q=0$, they reduce to that of the Schwarzschild metric \cite{DeFalco20183D,Bakala2019}.

\subsection{Test particle-radiation field interaction}
\label{sec:radfield}
We assume that the radiation-test particle interaction occurs through Thomson scattering, characterized by a constant momentum-transfer cross section $\sigma$, independent from direction and frequency of the radiation field. The radiation force is \cite{Bini2009,Bini2011,DeFalco20183D,Bakala2019,Defalco2020III}
\begin{equation} \label{radforce}
{\mathcal F}_{\rm (rad)}(\boldsymbol{U})^{\hat\alpha} = -\sigma P(\boldsymbol{U})^{\hat\alpha}{}_{\hat\beta} \, T^{\hat\beta}{}_{\hat\mu} \, U^{\hat\mu} \,,
\end{equation}
where $P(\boldsymbol{U})^{\hat\alpha}{}_{\hat\beta}=\delta^{\hat\alpha}_{\hat\beta}+U^{\hat\alpha} U_{\hat\beta}$ projects a vector orthogonally to $\boldsymbol{U}$. Decomposing the photon four-momentum $\boldsymbol{k}$ first with respect to the test particle four-velocity, $\boldsymbol{U}$, and then in the local observer frame, $\boldsymbol{n}$, we have \cite{DeFalco20183D}
\begin{equation} \label{diff_obg}
\boldsymbol{k} = E(\boldsymbol{n})[\boldsymbol{n}+\boldsymbol{\hat{\nu}}(\boldsymbol{k},\boldsymbol{n})]=E(\boldsymbol{U})[\bold{U}+\boldsymbol{\hat {\mathcal V}}(\boldsymbol{k},\boldsymbol{U})].
\end{equation}
Exploiting Eq. (\ref{diff_obg}) in Eq. (\ref{radforce}), we obtain \cite{DeFalco20183D,Bakala2019,Defalco2020III}
\begin{equation} \label{Frad0}
\begin{aligned}
{\mathcal F}_{\rm (rad)}(\boldsymbol{U})^{\hat\alpha}&=-\sigma \Phi^2 [P(\boldsymbol{U})^{\hat\alpha}{}_{\hat\beta} k^{\hat\beta}]\, (k_{\hat\mu} U^{\hat\mu})\\
&=\sigma \, [\Phi E(\boldsymbol{U})]^2\, \hat {\mathcal V}(\boldsymbol{k},\boldsymbol{U})^{\hat\alpha}\,.
\end{aligned}
\end{equation}
The equations of motion are $m \bold{a}(\boldsymbol{U}) = \boldsymbol{{\mathcal F}_{\rm (rad)}}(\boldsymbol{U})$, where $m$ is the test particle mass. Defined $\tilde \sigma=\sigma/m$, we obtain the following equations \cite{DeFalco20183D,Bakala2019,Defalco2020III}
\begin{equation}\label{geom}
\bold{a}(\boldsymbol{U})=\tilde \sigma \Phi^2 E(\boldsymbol{U})^2  \,\boldsymbol{\hat {\mathcal V}}(\boldsymbol{k},\boldsymbol{U}).
\end{equation}
Multiplying scalarly Eq. (\ref{diff_obg}) by $\boldsymbol{U}$, we find \cite{DeFalco20183D,Bakala2019,Defalco2020III}
\begin{equation} \label{enepart}
E(\boldsymbol{U})=\gamma E(\boldsymbol{n})[1-\nu\sin\psi\cos(\alpha-\beta)].
\end{equation}
Such splitting permits to determine $\boldsymbol{\hat{\mathcal{V}}}(\boldsymbol{k},\boldsymbol{U})=\hat{\mathcal{V}}^t\boldsymbol{n}+\hat{\mathcal{V}}^r\boldsymbol{e_{\hat r}}+\hat{\mathcal{V}}^\theta \boldsymbol{e_{\hat\theta}}+\hat{\mathcal{V}}^\varphi \boldsymbol{e_{\hat\varphi}}$ as \cite{DeFalco20183D,Bakala2019,Defalco2020III}
\begin{eqnarray}
&&\hat{\mathcal{V}}^{\hat r}=\frac{\sin\beta}{\gamma [1-\nu\sin\psi\cos(\alpha-\beta)]}-\gamma\nu\sin\psi\sin\alpha,\label{rad1}\\
&&\hat{\mathcal{V}}^{\hat \theta}=-\gamma\nu\cos\psi \label{rad2},\\
&&\hat{\mathcal{V}}^{\hat\varphi}=\frac{\cos\beta}{\gamma [1-\nu\sin\psi\cos(\alpha-\beta)]}-\gamma\nu\sin\psi\cos\alpha,\label{rad3}\\
&&\hat{\mathcal{V}}^{\hat t}=\gamma\nu\left[\frac{\sin\psi\cos(\alpha-\beta)-\nu}{1-\nu\sin\psi\cos(\alpha-\beta)}\right].\label{rad4}
\end{eqnarray}

\subsection{Equations of motion}
\label{sec:eqm}
The test particle equations of motion are written in terms of magnitude of spatial velocity $\nu$, polar  $\psi$ and azimuthal $\alpha$ angles of the spatial velocity measured in the local ZAMO frame, radius $r$, polar angle $\theta$, and independent from the azimuthal angle $\varphi$ due to rotational symmetry of the PR model \cite{DeFalco20183D,Bakala2019,Defalco2020III},
\begin{eqnarray}
\frac{d\nu}{d\tau}&=& -\frac{1}{\gamma}\bigg\{\sin\alpha \sin\psi\left[a(\boldsymbol{n})^{\hat r}+2\nu\cos\alpha\sin\psi\theta(\boldsymbol{n})^{\hat r}{}_{\hat \varphi}\right]\notag\\
&& +\cos\psi\left[a(\boldsymbol{n})^{\hat \theta}+2\nu\cos\alpha\sin\psi\theta(\boldsymbol{n})^{\hat \theta}{}_{\hat \varphi}\right]\bigg\}\label{EoM1}\\
&&+\frac{\tilde{\sigma}[\Phi E(\boldsymbol{U})]^2}{\gamma^3\nu}\hat{\mathcal{V}}^{\hat t},\nonumber\\
\frac{d\psi}{d\tau}&=& \frac{\gamma}{\nu} \left\{-\sin\alpha\cos\psi \left[a(\boldsymbol{n})^{\hat r}+2\nu\cos\alpha\sin\psi\theta(\boldsymbol{n})^{\hat r}{}_{\hat \varphi}\right]\right.\label{EoM2}\notag\\
&&\left.+\sin\psi \left[a(\boldsymbol{n})^{\hat \theta}+2\nu\cos\alpha\sin\psi\theta(\boldsymbol{n})^{\hat \theta}{}_{\hat \varphi}\right]\right.\\
&&\left.+\nu^2\left[\left(k(\varphi,\boldsymbol{n})^{\hat \theta}\cos^2\alpha+k(r,\boldsymbol{n})^{\hat \theta}\sin^2\alpha\right)\sin\psi\right.\right.\nonumber\\
&&\left.\left.-k(\theta,\boldsymbol{n})^{\hat r}\sin\alpha\cos\psi\right]\right\}\nonumber\\
&&+\frac{\tilde{\sigma}[\Phi E(\boldsymbol{U})]^2}{\gamma\nu^2\sin\psi}\left[\hat{\mathcal{V}}^{\hat t}\cos\psi-\hat{\mathcal{V}}^{\hat \theta}\nu\right],\nonumber\\
\frac{d\alpha}{d\tau}&=&\frac{\gamma\cos\alpha}{\nu\sin\psi}\left\{-\left[a(\boldsymbol{n})^{\hat r}+2\nu\cos\alpha\sin\psi\theta(\boldsymbol{n})^{\hat r}{}_{\hat \varphi}\right]\right.\label{EoM3}\\
&&\left.-\nu^2\left[\left(k(\varphi,\boldsymbol{n})^{\hat \theta}-k(r,\boldsymbol{n})^{\hat \theta}\right)\cos\psi\sin\psi\sin\alpha\right.\right.\notag\\
&&\left.\left.+k(\varphi,\boldsymbol{n})^{\hat r}\sin^2\psi+k(\theta,\boldsymbol{n})^{\hat r}\cos^2\psi\right]\right\}\nonumber\\
&&+\frac{\tilde{\sigma}[\Phi E(\boldsymbol{U})]^2\cos\alpha}{\gamma\nu\sin\psi}\left[\hat{\mathcal{V}}^{\hat r}-\hat{\mathcal{V}}^{\hat \varphi}\tan\alpha\right],\nonumber\\
U^r&\equiv&\frac{dr}{d\tau}=\frac{\gamma\nu\sin\alpha\sin\psi}{\sqrt{g_{rr}}}, \label{EoM4}\\
U^\theta&\equiv&\frac{d\theta}{d\tau}=\frac{\gamma\nu\cos\psi}{\sqrt{g_{\theta\theta}}} \label{EoM5},\\
U^\varphi&\equiv&\frac{d\varphi}{d\tau}=\frac{\gamma\nu\cos\alpha\sin\psi}{\sqrt{g_{\varphi\varphi}}}-\gamma \frac{N^\varphi}{N}.\label{EoM6}
\end{eqnarray}

Defining $A=\tilde{\sigma}\Phi_0^2E^2$, which is the so-called luminosity parameter, and can be also written as $A/M=L/L_{\rm EDD}\in[0,1]$, where $L$ is the luminosity measured by a static observer at infinity, and $L_{\rm EDD}$ is the Eddington luminosity \cite{DeFalco20183D,Bakala2019,Defalco2020III}. Using Eqs. (\ref{INT_PAR}) and (\ref{enepart}), we obtain
\begin{equation}
\tilde{\sigma}[\Phi E(\boldsymbol{U})]^2=\frac{A\gamma^2 (1+bN^\varphi)[1-\nu\sin\psi\cos(\alpha-\beta)]^2}{N^2 r^2F_3(r,\theta)\sin\beta}.
\end{equation}

\section{Critical hypersurfaces}
\label{sec:CH}
The dynamical system governed by Eqs. (\ref{EoM1}) -- (\ref{EoM6}) admits, as the previous models, a critical hypersurface outside of the emitting surface, where gravitational attraction and radiation pressure balance. Such region is analytically determined by the critical radius $r_{\rm crit}$ as function of $\theta$, i.e., $r_{\rm crit}=r_{\rm crit}(\theta)$, once the parameters $(J,Q,A,R_\star,\Omega_\star)$ are assigned. We consider a test particle moving along a non-equatorial plane on purely circular orbit (i.e., the azimuthal and polar angles related to the test particle' spatial velocity as measured in the local ZAMO frame are respectively $\alpha=0,\pi$, $\psi=\pi/2$, and the magnitude of the spatial velocity is $\nu=\mbox{const}$). Equation (\ref{EoM1})
for $d\nu/d\tau=0$ reduces to \cite{Bini2009,Bini2011,DeFalco20183D,Bakala2019,Defalco2020III}
\begin{equation} 
\tilde{\sigma}[\Phi E(\boldsymbol{U})]^2\hat{\mathcal{V}}^{\hat t}=0,\quad \Rightarrow\quad \nu=\cos\beta.
\end{equation}
The velocity of the test particle equates the photon azimuthal velocity. Since the test particle moves tangentially on the critical hypersurface, we have $d\alpha/d\tau=0$, and Eq. (\ref{EoM3}) assumes the following form
\begin{equation} \label{eq:CH}
\begin{aligned}
&a(\boldsymbol{n})^{\hat r}+2\nu\theta(\boldsymbol{n})^{\hat r}{}_{\hat \varphi}+\nu^2k(\varphi,\boldsymbol{n})^{\hat r}\\
&=\frac{\tilde{\sigma}[\Phi E(\boldsymbol{U})]^2}{\gamma^2}\hat{\mathcal{V}}^{\hat r},
\end{aligned}
\end{equation}
which is an implicit equation for determining the critical radius $r_{\rm crit}$ \cite{Bini2009,Bini2011,DeFalco20183D,Bakala2019,Defalco2020III}. The critical hypersurface is axially symmetric with respect to the polar direction, and can assume either an oblate or prolate form depending on the interplay between gravitational pull $a(\boldsymbol{n})^{\hat r}$, centrifugal forces $k(\varphi,\boldsymbol{n})^{\hat r}$, frame dragging effect $2\nu\theta(\boldsymbol{n})^{\hat r}{}_{\hat \varphi}$, and radiation forces including the PR effect $\frac{\tilde{\sigma}[\Phi E(\boldsymbol{U})]^2}{\gamma^2}\hat{\mathcal{V}}^{\hat r}$.

\begin{figure*}[th!]
	\centering
\vbox{
	\hbox{
		\includegraphics[scale=0.24]{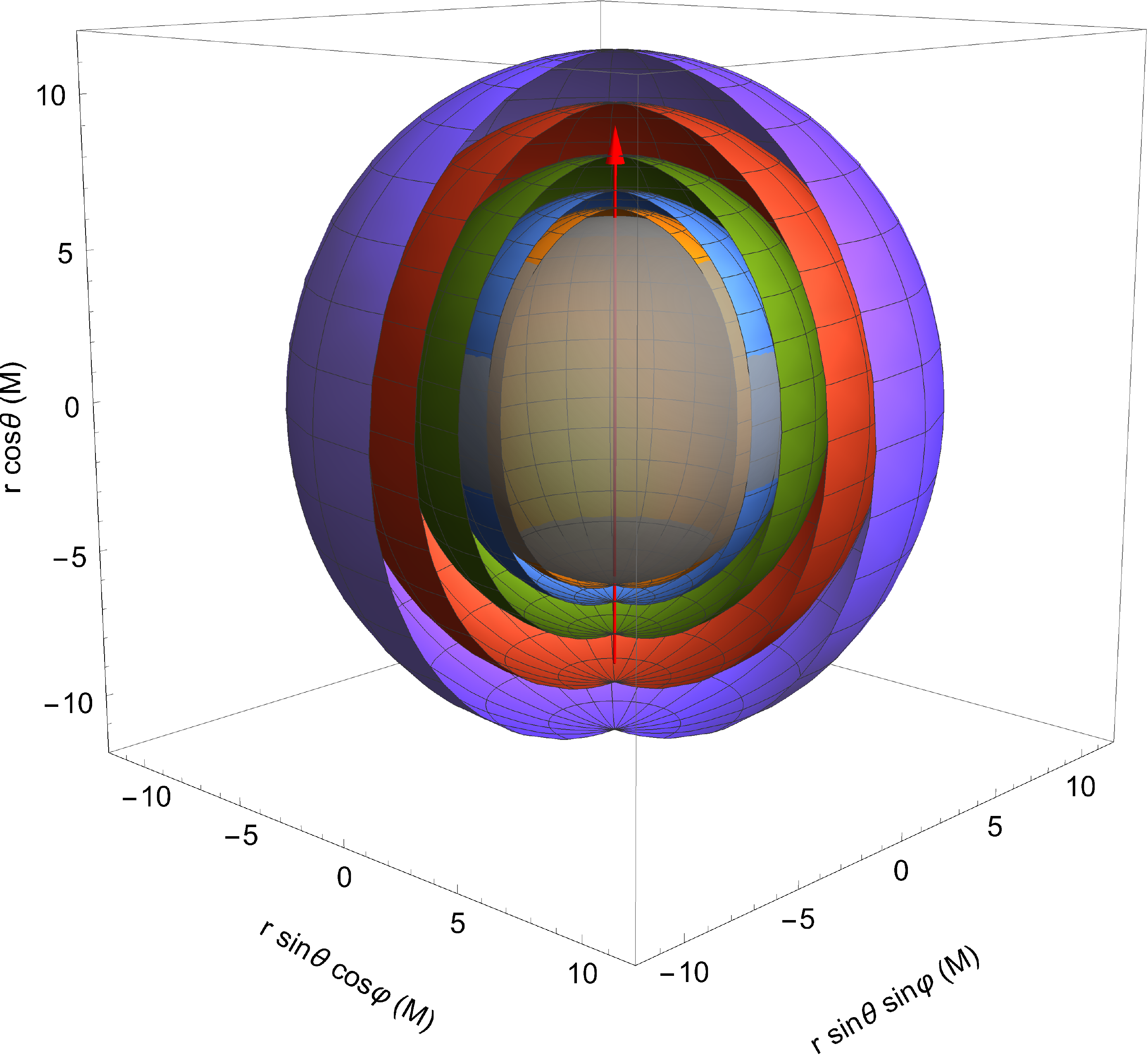}
		\hspace{0.5cm}
		\includegraphics[scale=0.34]{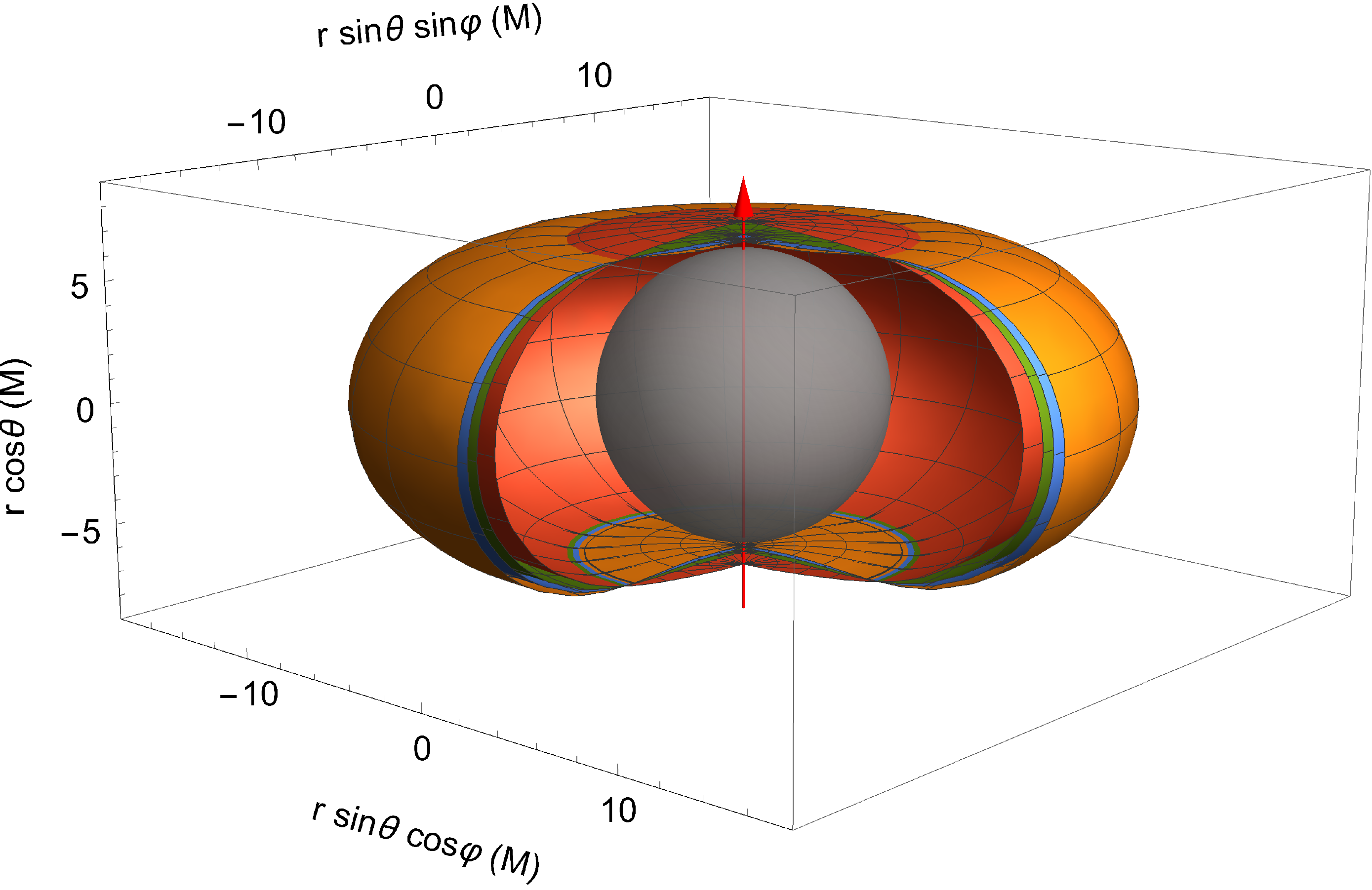}}
\vspace{0.3cm}		
	\hbox{
		\includegraphics[scale=0.31]{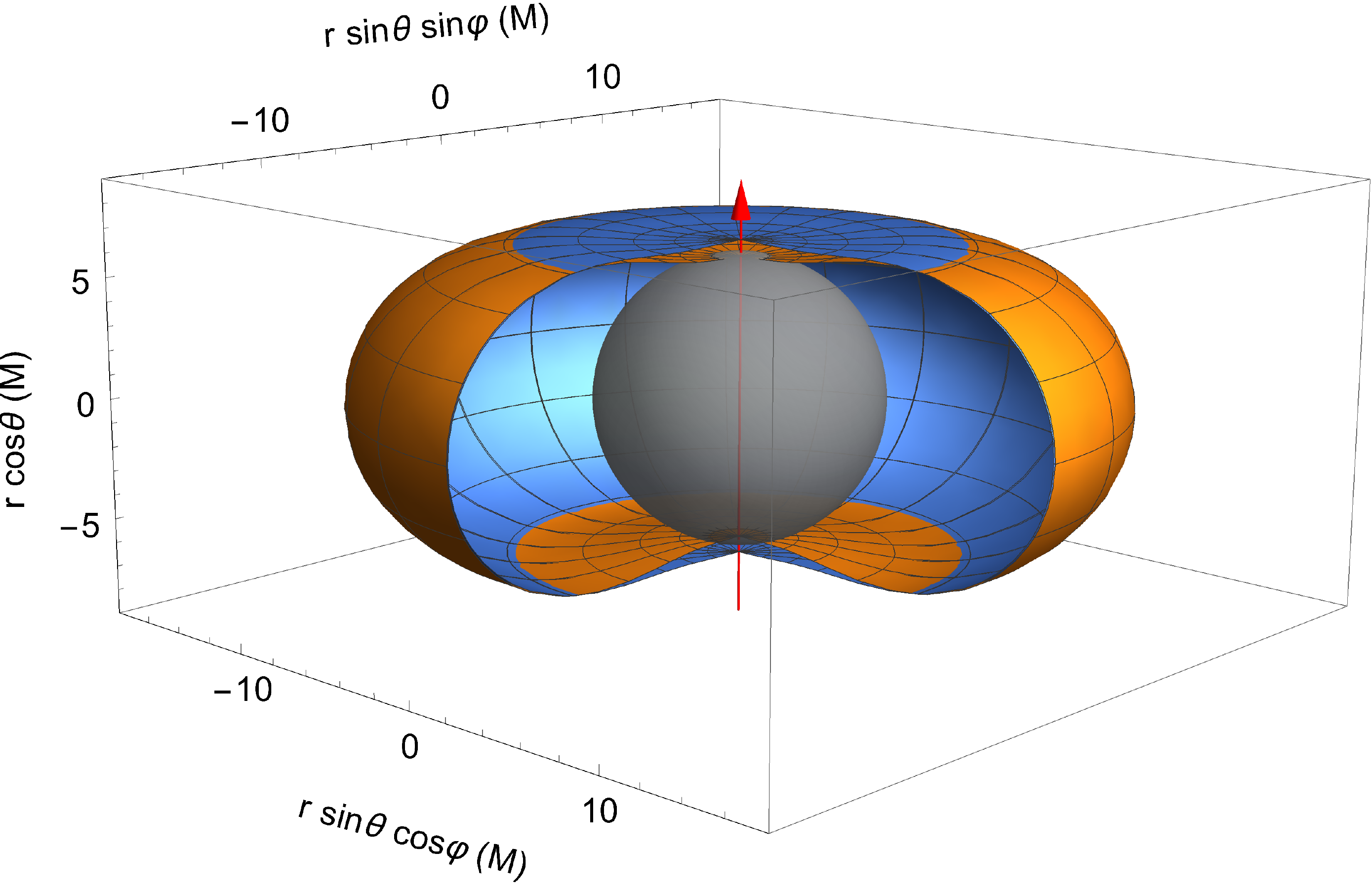}
		\hspace{0.3cm}
		\includegraphics[scale=0.31]{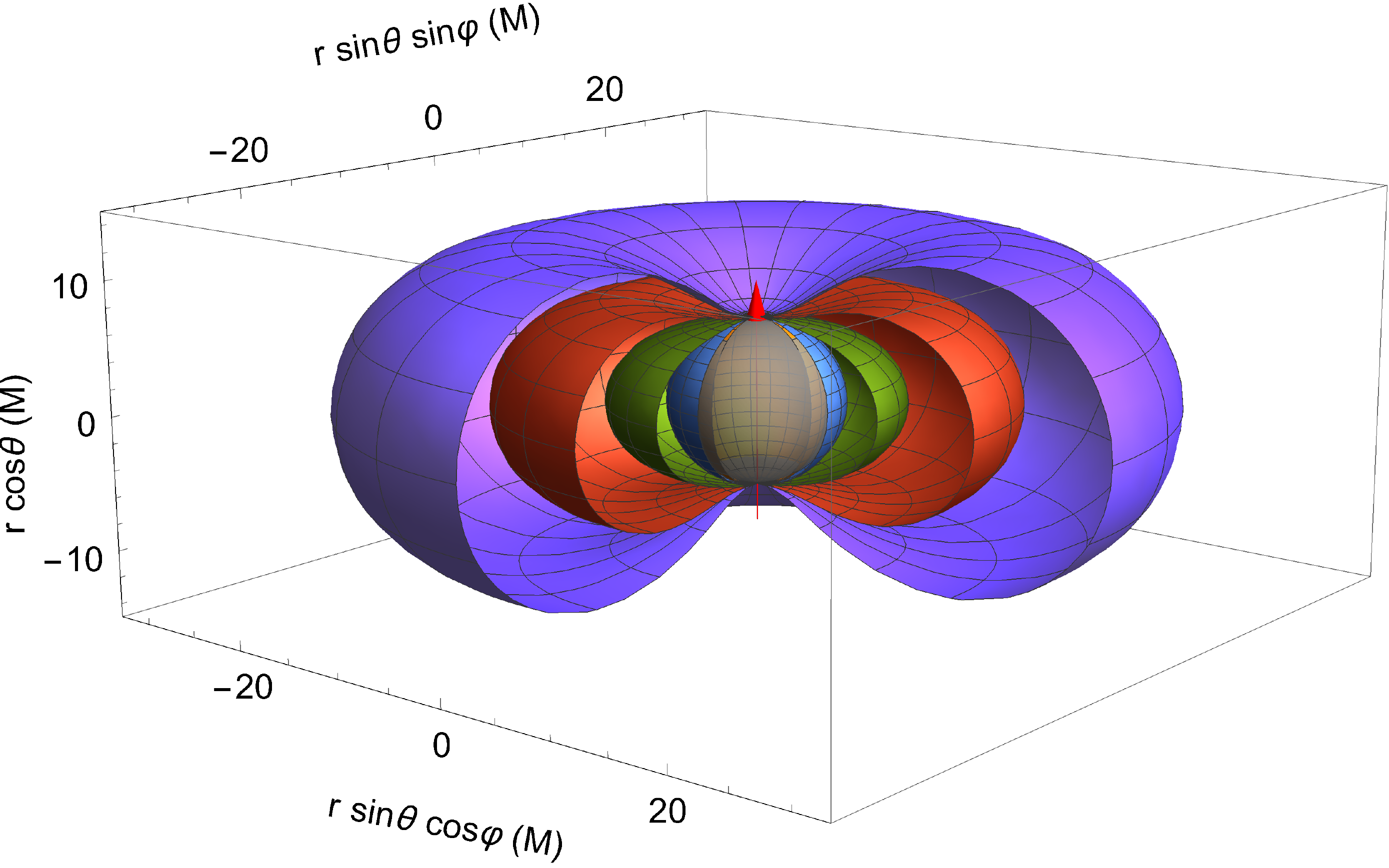}}} 		
	\caption{We show in different plots the shapes of the critical hypersurfaces in terms of varying parameters. In all plots the emitting surface radius is $R_\star=6M$ and is displayed as a gray surface, and the red arrow points in the positive polar direction. 
Left upper panel: We fix $a=0.1$, $q=-0.3$, $\Omega=69\times10^{-4}M^{-1}$ and change the relative luminosity $A=0.80,\, 0.82, \,0.85, \,0.88, \,0.90$. 
Right upper panel: We fix $A=0.8$, $q=-0.3$, $\Omega=0.034 M^{-1}$ and change the spin $a=0.1,\, 0.2, \,0.3,\,0.5$. 
Left lower panel: We fix $A=0.8$, $a=0.1$, $\Omega=0.034 M^{-1}$ and change the quadrupole moment $q=-0.1,\,-0.4$. 
Right lower panel: We fix $A=0.8$, $a=0.1$, $q=-0.3$, and change the critical hypersurface angular velocity $\Omega=0.019 M^{-1}, 0.029 M^{-1}, 0.038 M^{-1}, 0.048 M^{-1}$. }
	\label{fig:Fig1}
\end{figure*}

In Fig. \ref{fig:Fig1}, we plot different configurations of critical hypersurfaces by varying the values of some parameters. 
Bearing in mind the condition (\ref{eq:HTvalidity}), which tells from which radius the Hartle-Thorne metric is valid, we decided to display also the \emph{unphysical solutions}, which are located inside the emitting source (gray surface) or not respecting the above requirement, because we would like to highlight how the critical hypersurface configurations morph in terms of the parameters' variability.
For convenience we have defined the Hartle-Thorne spin $a\equiv cJ/(GM^2)\in[0,1]$ and quadrupole moment $q\equiv -c^4Q/(G^2M^3)\in[-1,0]$. It is also important to note that since the Hartle-Thorne metric is an approximate solution in terms of $a$ and $q$, it works for $a,-q\ll1$. We use only negative values of the quadrupole parameter $q$, otherwise we checked that no critical hypersurface exists. A physical explanation of the occurrence of such a phenomenon can be attributed to the combined effect of the centrifugal and frame-dragging forces, which are responsible for sweeping the test particle away.

We immediately see that the luminosity parameter $A$ plays a fundamental role in shaping the critical hypersurface. In particular, high luminosities $A=L/L_{\rm Edd}\gtrsim0.7$ are needed to have a critical hypersurface relatively far from the emitting surface (see upper left panel). Increasing the spin values, the critical hypersurface becomes more oblate (see upper right panel), and the same argument holds also for the quadrupole moment (see lower left panel). The rotation of the emitting surface, $\Omega_\star$, strongly contributes also in shaping the form of the critical hypersurface (see lower right panel). In conclusion, we can infer that the radiation field intensity and the different gravitational effects (i.e., curved geometry, frame dragging, and centrifugal forces) only along the radial direction contribute to morph the critical hypersurface. 

In the next sections, we derive the conditions to obtain suspended orbits (Sec. \ref{sec:SO}), and we apply this model to describe the emission properties of a NS (Sec. \ref{sec:NS}).

\subsection{Suspended orbits}
\label{sec:SO}
The test particle could move on circular orbits bounded on the critical hypersurface at constant height $y\neq0$ (off-equatorial plane), without the action of the latitudinal drift mechanism (see Refs. \cite{DeFalco20183D,Bakala2019}, for further details). To obtain such configurations, the test particle must touch the critical hypersurface with the following conditions: $\alpha=0,\pi$, $\nu=\cos\beta$, $r=r_{\rm crit}(\theta)$, and $d\psi/d\tau=0$ (where this last condition is the strong constraint for not having latitudinal drift towards the equatorial plane). Vanishing Eq. (\ref{EoM2}), it is possible to determine the value of $\psi$, by solving this implicit equation \cite{Bakala2019}:
\begin{equation}\label{eq:SO}
\begin{aligned}
&a(\boldsymbol{n})^{\hat \theta}+2\nu\sin\psi\theta(\boldsymbol{n})^{\hat \theta}{}_{\hat \varphi}+\nu^2k(\varphi,\boldsymbol{n})^{\hat \theta}\\
&+\frac{\tilde{\sigma}[\Phi E(\boldsymbol{U})]^2}{\gamma^2 \nu\sin^2\psi}\left[\hat{\mathcal{V}}^{\hat t}\cos\psi-\hat{\mathcal{V}}^{\hat \theta}\nu\right]=0.
\end{aligned}
\end{equation}
The value of $\psi$ strongly depends on emitting surface location $R_\star$, angular velocity $\Omega_\star$, and compact object quadrupole moment $q$ and spin $a$. 
\begin{figure*}[th!]
	\centering
	\hbox{
		\includegraphics[scale=0.3]{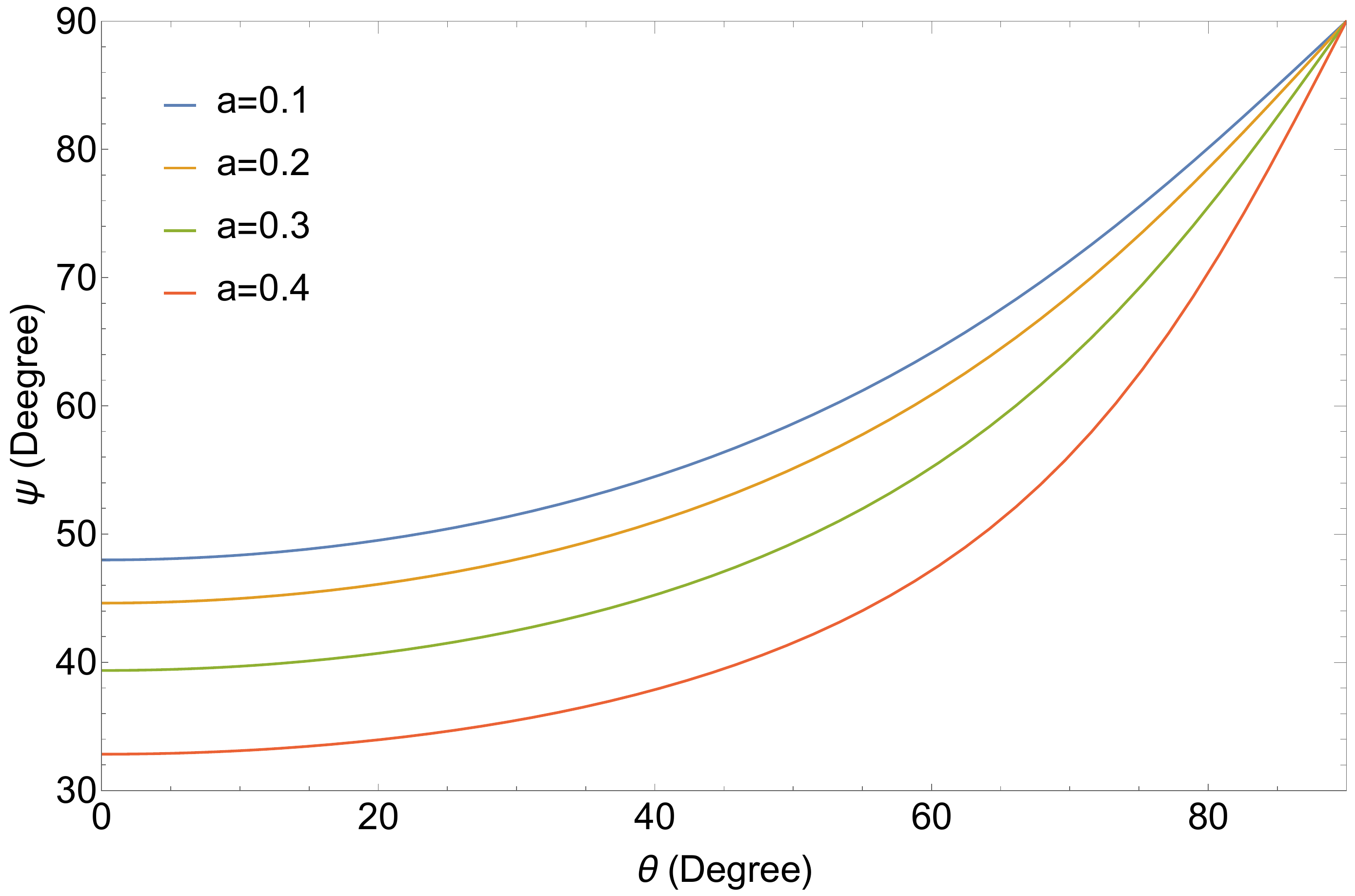}
		\hspace{0.3cm}
		\includegraphics[scale=0.3]{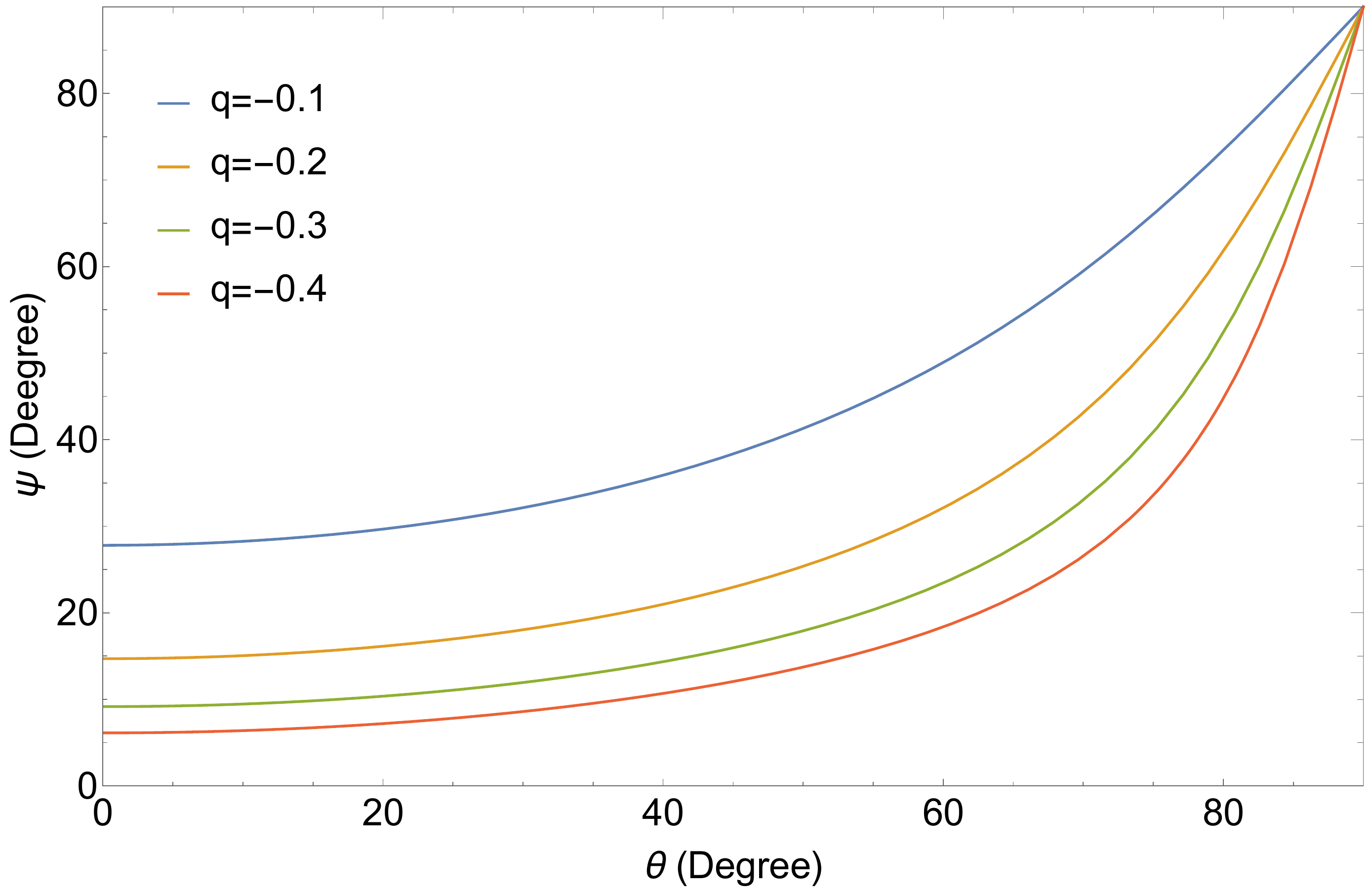}}
	\caption{Different configurations of suspended orbits at height $\theta$ on the critical hypersurface in terms of the angle $\psi$ are shown. We set $A=0.8$, $\Omega_\star=0.01$, and $R_\star=3M$. Once we fix $q=-0.4$, and change the value of the spin $a=0.1,0.2,0.3,0.4$ (see left panel), and then we fix $a=0.1$ and change the value of the quadrupole moment $q=-0.1,-0.2,-0.3,-0.4$ (see right panel).}
	\label{fig:Fig2}
\end{figure*}

In Fig. \ref{fig:Fig2} we show the angle $\psi$ at which the test particle should touch the critical hypersurface to reach the fixed height $\theta$, and moving on such plane on a circular orbit. We note that it is always possible to have suspended orbits both on and off the equatorial plane. It is interesting to note that increasing the module of the spin the $\psi$ angle decreases because the test particle has to contrast stronger forces (see left panel). The same argument holds also for the module of $q$ (see right panel). These particular configurations, on which a test particle moves stably, are typical of the general relativistic PR effect in the 3D space \cite{DeFalco20183D,Bakala2019,Defalco2020III}. Their formation is a result of the perfect balance among the gravitational contributions (including also the polar centrifugal force and the frame-dragging effect) and the radiation forces along the polar direction, see Eq. (\ref{eq:SO}).

We note that off-equatorial suspended orbits are the consequence of our assumptions on the radiation stress-energy tensor (\ref{eq:radfield}), which is constituted by \qm{a single stream of photons} reaching the test particle at each instant of time. In other models of the 3D general relativistic PR effect \cite{Wielgus2019}, the radiation source is modeled by a bunch of photons coming from the whole 3D emitting surface. In this case, the radiation force drives always the test particle toward the equatorial plane, where it moves stably, since there is a perfect balance of both gravitational and radiation forces from the two hemispheres of the emitting surface.

\subsection{Radiation effects around a neutron star}
\label{sec:NS}
The present model of the general relativistic PR effect in the Hartle-Thorne metric can be used to describe several radiation processes occurring on and around a NS, like: accretion phenomena, type-I X-ray bursts, photospheric radius expansion. To see how to apply our developments to a NS, we know that such an astrophysical object is described by mass $M$, radius $R_\star$, spin frequency $f$ (or angular velocity $\Omega_\star$), Hartle-Thorne angular momentum $a$ and quadrupole moment $q$. Since we have a non-spherical distribution of the mass, we can also consider that the NS shape is not anymore spherical, but deformed as an ellipsoid. However, due to the axially symmetry of the Hartle-Thorne spacetime, it is reasonable to assume that this ellipsoid is rotationally symmetric and therefore it is defined by the equatorial $R_{\rm eq}$ and polar $R_{\rm pol}$ radii. Therefore, the NS form is described by
\begin{equation}
\frac{x^2+y^2}{R_{\rm eq}^2}+\frac{z^2}{R_{\rm pol}^2}=1,\ \ \Leftrightarrow\ \
\begin{cases}
x=R_{\rm eq}\sin\theta\cos\varphi,\\
y=R_{\rm eq}\sin\theta\sin\varphi,\\
z=R_{\rm pol}\cos\theta.
\end{cases}
\end{equation}
The polar radius can be also expressed in terms of the ellipticity $e$, namely $R_{\rm pol}=R_{\rm eq}(1+e)$.

Therefore, the NS is defined by six parameters $\left\{M,f,R_{\rm eq},a,q,e\right\}$. Since we have already several other parameters for characterizing the radiation processes, we would like to reduce the NS parameter space. To this end, we follow the approach of Baub{\"o}ck and collaborators \cite{Baubock2013}, defining the following set of parameters:
\begin{equation} 
\begin{aligned}
&f_0=\frac{1}{2\pi}\sqrt{\frac{GM}{R_{\rm eq}^3}}, \quad \epsilon_0=\frac{f}{f_0},\quad a=\epsilon_0 a^*,\\
& q=\epsilon_0^2 q^*,\quad e=\epsilon_0^2 e^*,\quad \zeta=\frac{GM}{c^2R_{\rm eq}},
\end{aligned}
\end{equation}
where $f_0$ is the Keplerian angular velocity of a test particle orbiting at a radius $R_{\rm eq}$ around a mass $M$, corresponding also to the maximum NS frequency to which it can be spun up before breakup. We label the parameters with asterisk in order to highlight that they depend on the particular NS equation of state considered. 

Baub{\"o}ck and collaborators show that the actual six-parameter space can be reduced to a three-parameter space spanned by $\left\{M, R_{\rm eq}, f\right\}$, which is valid over the astrophysically relevant parameter range and for a variety of equations of state. Therefore, the remaining three parameters $\left\{a,q,e\right\}$ can be written in terms of $\left\{M, R_{\rm eq}, f\right\}$ through the following equations \cite{Baubock2013}:
\begin{eqnarray} 
a&=&\epsilon_0(1.1035-2.146\zeta+4.5756\zeta^2), \label{eqs:NS1}\\
q&=&a^2\exp\left[-2.014+0.601\log\left(\frac{a}{\epsilon_0}\zeta^{-3/2}\right)\right.\label{eqs:NS2}\\
&&\left.+1.10\log\left(\frac{a}{\epsilon_0}\zeta^{-3/2}\right)^2-0.412\log\left(\frac{a}{\epsilon_0}\zeta^{-3/2}\right)^3\right.\notag\\
&&\left.+0.0459\log\left(\frac{a}{\epsilon_0}\zeta^{-3/2}\right)^4\right],\notag\\
e&=&\frac{\epsilon_0^2}{32\zeta^3}\left\{2\zeta\left[8\zeta^2-32a^*\zeta^{7/2}+8a^*{}^2\zeta^5-48a^*{}^2\zeta^6\right.\right.\label{eqs:NS3}\\
&&\left.\left.+(a^*{}^2-q^*)(45-135\zeta+60\zeta^2+30\zeta^3)+24a^*{}^2\zeta^4\right]\right.\notag\\
&&\left.+45(a^*{}^2-q^*)(1-2\zeta)^2\log(1-2\zeta)\right\}.\notag
\end{eqnarray}

\begin{figure}[th!]
	\centering
	\includegraphics[scale=0.3]{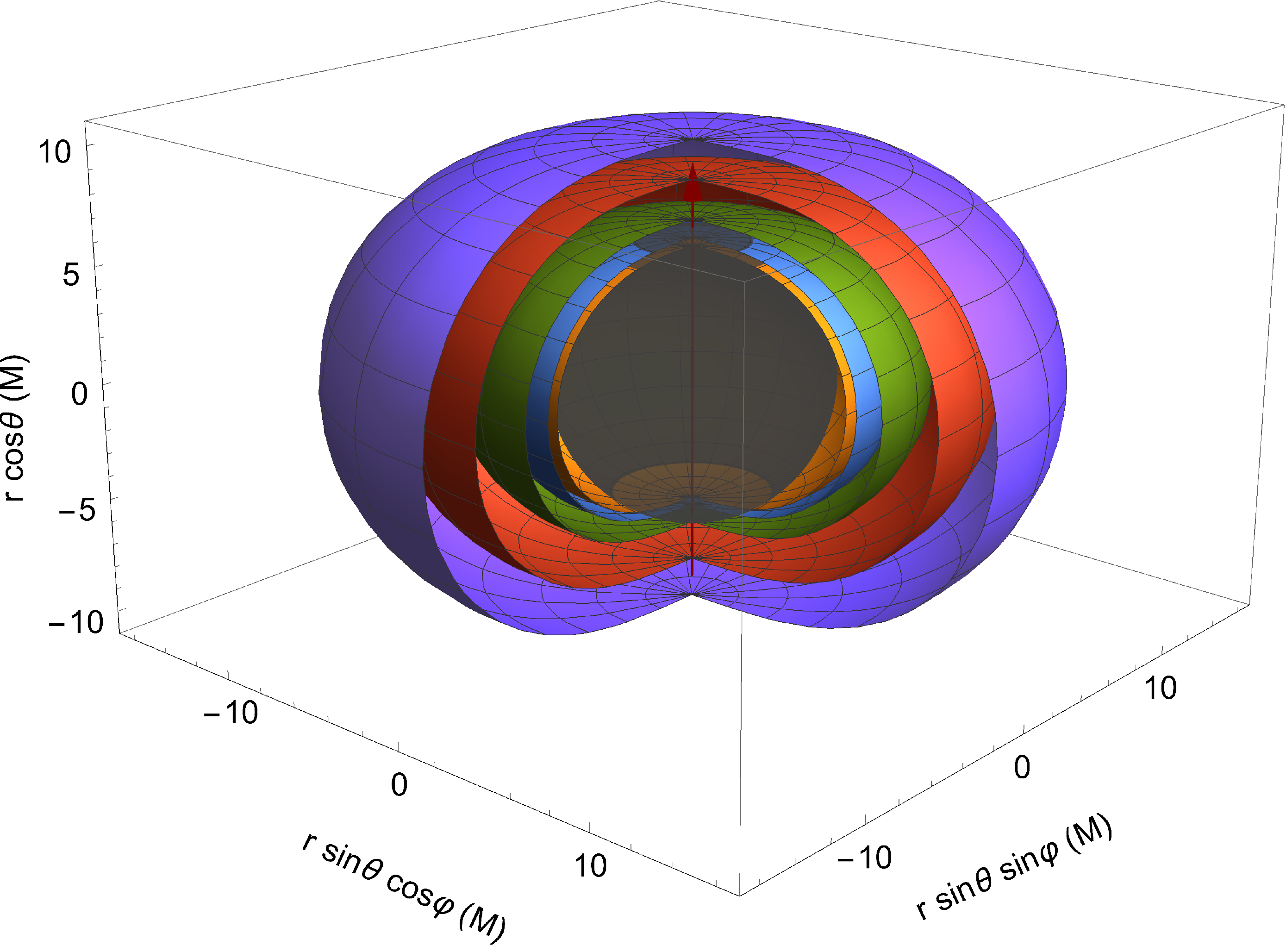}
	\caption{Critical hypersurfaces for a NS of mass $M=1.4M_\odot$, equatorial radius $R_{\rm eq}=6M$, and spin frequency $f=400$ Hz, and for different luminosity parameter values $A=0.82,0.83,0.85,0.88,0.9$ (corresponding respectively to the orange, blue, green, red, and violet colors of the displayed surfaces). The gray surface represents the NS and the red arrow the positive polar direction.}
	\label{fig:Fig4}
\end{figure}

We note that since we assumed an ellipsoid shape, where the NS radius can be written as 
\begin{equation}
R_{\rm NS}(\theta)=\sqrt{R_{\rm eq}^2\sin^2\theta+R_{\rm pol}^2\cos^2\theta},
\end{equation}
we have that the formula to calculate the impact parameter (\ref{eq:imp_para}) must be slightly changed in the following form
\begin{equation} \label{eq:imp_para2}
b=\left[-\frac{g_{t\varphi}+g_{\varphi\varphi}\Omega_\star}{g_{tt}+g_{t\varphi}\Omega_\star}\right]_{r=R_{\rm NS}(\theta)}.
\end{equation}
We can also relate the angular velocity of the emitting surface $\Omega_\star$ in terms of the frequency $f$ through 
\begin{equation}
\Omega_\star=2\pi f\frac{GM}{c^3}.
\end{equation}
To further reduce our parameter space, we fix the values of the NS mass $M=1.4M_\odot$, and equatorial radius $R_{\rm eq}=6M$, therefore the remaining free parameter is only $f$. In Fig. \ref{fig:Fig4}, we plot different NS critical hypersurfaces for different values of the luminosity parameter $A$. We have checked also how the critical hypersurfaces would have altered its shape, if we had varied $f\in[0,400]$ Hz (physically allowed NS frequency range), but no significant changes have been found. Therefore, in Fig. \ref{fig:Fig4} we fix $f=400$ Hz. In such cases, we have the following values of the dependent parameter set: $R_{\rm pol}=6.23M$, $\Omega_\star=17\times10^{-3}\ M^{-1}$ , $a=0.22$, $q=0.29$, and $e=0.04$. This example confirms that even if we consider a high spin frequency, $f = 400$ Hz, the above approach
results in a rather low spin parameter, $a = 0.22$, within the applicability of the Hartle-Thorne spacetime model.

\begin{figure}[th!]
	\centering
	\includegraphics[scale=0.3]{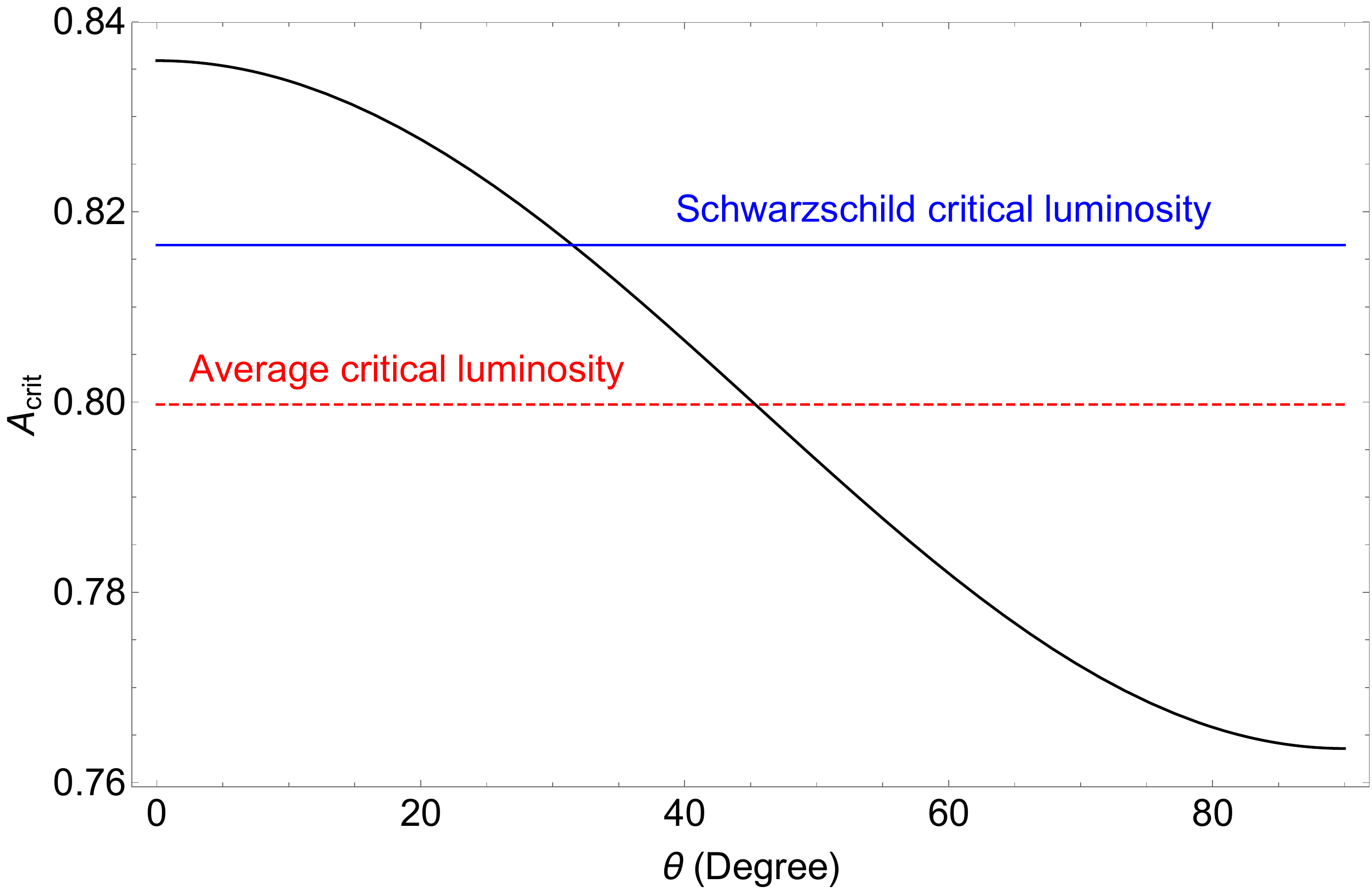}
	\caption{Critical luminosity $A_{\rm crit}$ in terms of the polar angle $\theta$ for $M=1.4M_\odot$, $R_{\rm eq}=6M$, $f=400$ Hz. The red dashed line represents the average critical luminosity, while the blue continuous line is the critical luminosity for $f=0$ (i.e., $a=0$ and $q=0$), namely it is framed in the Schwarzschild spacetime.}
	\label{fig:Fig5}
\end{figure}

From this plot we also note that for having critical hypersurfaces outside of the NS surface, we need to have quite high-luminosities, namely $A=L/L_{\rm Edd}\gtrsim0.82$. In Fig. \ref{fig:Fig5} we determine the critical luminosity $A_{\rm crit}$, being the luminosity $A$ at which the critical hypersurfaces touches the NS surface at height $\theta$. The critical luminosity at the equatorial plane is $A_{\rm crit}=0.7654$, and at the poles is $A_{\rm crit}=0.836$; while the average critical luminosity is $A_{\rm crit}=0.800$, and finally the critical luminosity for $f=0$ (i.e., $a=0$ and $q=0$, the Schwarzschild metric), remains constant at $A_{\rm crit}=0.817$.

\section{Conclusions}
\label{sec:end}
We have developed, for the first time in the literature, the full general relativistic 3D (and 2D by considering $\theta=\psi=\pi/2$) treatment of the motion of a test particle around a non-spherical, and slowly rotating compact object, described by the Hartle-Thorne metric, and in the same time affected by the radiation field, including the general relativistic PR effect, from a spherical and rigidly rotating emitting surface located outside the compact object (see Sec. \ref{sec:dynamics}). The Hartle-Thorne spacetime is an approximate solution of the Einstein field equations in the vacuum, and it is described in terms of three parameters: the mass $M$,  the angular momentum $J$, and the quadrupole moment $Q$ (see Sec. \ref{sec:HT_met}). 

In order to make our approach more flexible in view of extension of this model for other metrics more realistic than the Hartle-Thorne description, we have casted our initial calculations in a modular form (see Sec. \ref{sec:ZAMOq}). Indeed, changing the functional form of the functions $F_1(r,\theta),F_2(r,\theta),F_3(r,\theta)$, and calculating the related derivatives with respect to $r$ and $\theta$, it is possible to obtain the ZAMO quantities and then to straightforwardly derive the equations of motion, the critical hypersurface, and the suspended orbits. In addition in the Schwarzschild limit (i.e., $q\to0$ and $a\to0$), we have that $F_1(r,\theta),F_2(r,\theta),F_3(r,\theta)\to1$, and the metric (\ref{eq:HTmetric}) reduces to the Schwarzschild spacetime. 

The critical hypersurface equation (\ref{eq:CH}) depends on the luminosity parameter $A=L/L_{\rm Edd}$, the photon impact parameter $b$ (or equivalently from radius $R_\star$ and angular velocity $\Omega_\star$ of the emitting surface), the spin parameter $a$, and quadrupole moment $q$, see Sec. \ref{sec:CH}. In Fig. \ref{fig:Fig1}, we have produced different configurations of the critical hypersurfaces by varying the values of the parameters. We have shown that the radiation field and the gravitational effects strongly contribute to morph the critical hypersurfaces. In addition, high luminosities $A\gtrsim0.7$ permits to have the PR critical hypersurfaces located outside the emitting surface, making them physically possible.

We have analysed also the suspended orbits, which are configurations where the test particle moves on bound off-equatorial circular orbits on the critical hypersurface at a given $\theta$ height, see Sec. \ref{sec:SO}. In Fig. \ref{fig:Fig2} we have plotted the angle $\psi$ at which the test particle should be sent in order to move on a suspended orbit, in terms of the height $\theta$, and once by fixing the quadrupole moment and changing the angular momentum (see left panel), and then viceversa (see right panel). Therefore, it is possible to obtain suspended orbits at all heights for different values of the parameters $a,q$, due to the perfect equilibrium between gravitational and radiation forces, without having latitudinal drift motion towards the equatorial plane.

Finally, we have also proposed an application of the PR effect to model NSs, see Sec. \ref{sec:NS}. In this case, we have found a way to write the parameters $\left\{q,a,e,\Omega_\star\right\}$ in terms of the quantities $\left\{M,R_{\rm eq},f\right\}$ through Eqs. (\ref{eqs:NS1}) -- (\ref{eqs:NS3}), which are valid for a variety of equations of state. We have considered that the NS is not anymore a spherical body, but it is an ellipsoid with rotational-azimuthal symmetry, and determined by the equatorial $R_{\rm eq}$ and polar $R_{\rm pol}$ radii, where the latter can be expressed also in terms of the eccentricity $e$. This entailed to slightly change the expression of the photon impact parameter $b$, see Eq. (\ref{eq:imp_para2}). We can further reduce the parameter space by setting $M=1.4M_\odot$ and $R_{\rm eq}=6M$ (typical NS mass and radius), so that all the parameters will depend only on the spin frequency $f$. In Fig. \ref{fig:Fig4} we plotted the critical hypersurfaces, but only by varying the luminosity parameter $A=L/L_{\rm Edd}$, because there are no significant change in terms of $f$. Then, we have also analysed at which luminosity $A=L/L_{\rm Edd}$ the critical hypersurface touches the NS surface at the height $\theta$, comparing also these configurations with the Schwarzschild case, see Fig. \ref{fig:Fig5}. This can be the initial set up for then developing astrophysical models involving radiation effects occurring either around or on the surface of a NS.

As future project, we aim at extending our treatment of the 3D general relativistic PR effect around a fast rotating and non-spherical quadrupolar massive source. These astrophysical objects can be modeled by several sophisticated metrics, which in general due to the complex treatment requires specific numerical treatments \cite{Friedman2013}. 

\section*{Acknowledgements}
The authors thank the anonymous referee for the useful remarks and comments given in the review process. V.D.F. thanks Gruppo Nazionale di Fisica Matematica of Istituto Nazionale di Alta Matematica for the support. M.W. acknowledges the support of the Black Hole Initiative at Harvard University, which is funded by grants from the John Templeton Foundation and the Gordon and Betty Moore Foundation to Harvard University.

\bibliography{references}

\end{document}